\documentclass[twocolumn]{IEEEtran}
\usepackage{authblk}
\usepackage{silence}
\usepackage[font=scriptsize]{caption}

\usepackage[pdftex]{graphicx}
\usepackage{amsmath, amsthm, amssymb}
\usepackage[table,xcdraw]{xcolor}
\usepackage{array}
\usepackage{cite}
\usepackage{amssymb}
\usepackage{color}
\usepackage{amsmath}
\usepackage{multirow}
\usepackage{dirtytalk}
\usepackage{soul}

\usepackage{algorithm} 
\usepackage{algpseudocode} 
\usepackage{enumerate}
\usepackage{fancyhdr}
\fancypagestyle{FirstPage}{

\lhead{This work is under review. Copyright may be transferred without notice, after which this version may no longer be accessible.}
\lfoot{Email: ussman@snu.ac.kr}

}

\begin{document}

\title{MEDS-Net: Multi-Encoder based Self-Distilled Network with Bidirectional Maximum Intensity Projections for Lung Nodule Detection}
\author[1,2]{Muhammad Usman}
\author[1,*]{Azka Rehman}
\author[1]{Abdullah Shahid}
\author[3]{Siddique Latif}
\author[1]{Shi‑Sub Byon}
\author[4]{Byoung-Dai Lee}
\author[1]{Sung‑Hyun Kim} 
\author[1]{Byung‑il Lee}
\author[2]{Yeong‑Gil Shin}
\affil[1]{Center for Artificial Intelligence in Medicine and Imaging, HealthHub Co. Ltd., Seoul, 06524, South Korea}
\affil[2]{Department of Computer Science and Engineering, Seoul National University, Seoul, 08826, Republic of Korea}
\affil[3]{University of Southern Queensland, Springfield, Queensland 4300, Australia}
\affil[4]{School of Computer Science and Engineering, Kyonggi University, Suwon, 16227, South Korea}
\affil[*]{Corresponding Author: azka@healthhub.kr}

\setcounter{Maxaffil}{0}
\maketitle

\begin{abstract}
Early detection of lung cancer is an effective way to improve the cure rate of patients. Therefore, it is crucial to accurately detect lung nodules in computed tomography (CT) images for early diagnosis of lung cancer. However, developing an accurate computer-aided detection (CADe) system for lung nodule detection is a challenging task owing to the complex lung's structure and the heterogeneity of the lung nodules. In this study, we propose a lung nodule detection scheme motivated by the clinical workflow of radiologists, which combines CT scan sub-volume with bidirectional maximum intensity projection (MIP) images in a single architecture. For this purpose, we designed a novel multi-encoder based network (MEDS-Net) that exploits self-distillation to effectively learn from three different types of inputs, including 3D sub-volume, forward, and backward MIP images. MEDS-Net exploits three different encoders to effectively extract the distinctive insights which are passed to the decoder block. Further, the decoder block leverages the self-distillation mechanism by connecting the distillation block, which contains four auxiliary lung nodule detectors. It helps to expedite the convergence and improves the learning ability of the proposed architecture. Finally, the proposed scheme reduces the false positives by complementing the main detector with auxiliary detectors. The proposed scheme has been rigorously evaluated on 888 scans of the LIDC-IDRI dataset and obtained a competition performance metric (CPM) score of 93.6\%. The results demonstrate that incorporating bidirectional MIP images enables MEDS-Net to accurately detect lung nodules and reduce the number of false positives using auxiliary detectors.
\end{abstract}

\section{Introduction}
\thispagestyle{FirstPage}

Lung cancer is the most frequent cancer type, with the highest mortality rates globally \cite{https://doi.org/10.3322/caac.21660}. The survival rate of
patients highly depends upon the early detection of pulmonary nodules \cite{baldwin2015prediction}. For diagnosing pulmonary nodules, computed tomography (CT) has been extensively employed and proven effective \cite{sluimer2006computer}. 
However, it is often a time-consuming and tedious job to manually identify nodules in CT scans. Radiologists need to read the CT scans slice by slice, and depending upon the slice thickness, a chest CT may contain hundreds of slices. Additionally, manual lung nodule detection is error-prone \cite{seemann1999usefulness}. To alleviate this problem, computer-aided detection (CADe) systems play an important role in aiding radiologists in accurately diagnosing lung cancer. 

Numerous efforts have been made in designing CADe systems; however, developing a highly accurate and robust CADe system is still challenging due to the heterogeneity of lung nodules and the high degree of similarity between the lung nodules and their surrounding tissues. Deep learning (DL) based techniques have made vast inroads in various medical imaging, including lung nodule detection \cite{zhang2018pulmonary,halder2020lung}.   
In DL-based CADe systems, convolutional neural networks (CNNs) have been extensively employed in 2D and 3D to achieve significantly improved performance \cite{monkam2019detection}. Such CADe system typically consists of two stages: screening and false positives (FPs) reduction. The former is typically applied to identify suspicious regions (i.e., nodule candidates) in a patient’s exam. The latter stage is to reduce the FPs by applying the additional CNNs\cite{cao2020two,pezeshk20183,zhao2022attentive, zhou2022cascaded}. Although such CADe systems demonstrated promising performance for lung nodule detection, these techniques add computational complexity by increasing the number of CNNs to train and are more prone to error as nodules missed at the first stage cannot be detected at the second stage. Subsequently, the sensitivity at the detection stage is kept high, significantly increasing the FPs and the load for the FPs reduction stage.

To overcome the challenges associated with dual-stage based CADe systems, a few studies proposed single-stage frameworks to detect the lung nodules \cite{li2020deepseed,zhu2022channel,luo2022scpm}. Such techniques considerably improve the computational complexities and ease the training and optimization process; however, they suffer from a low performance that limits their real-time clinical employment. The disconnectivity of existing CADe systems with clinical workflow is another reason for not being contentedly opted for by clinical practitioners. In the real-time clinical workflow, mostly, radiologists first examine the maximum intensity projection (MIP) images \cite{wallis1989three} of CT scan to roughly screen the nodule candidates and use the raw CT scan slices to finalize the decision. Only a few studies \cite{masood2020automated,zheng2019automatic} utilized the MIP images to improve lung nodule detection performance and comply with the clinical workflow. For instance, Masood \textit{et al.} \cite{masood2020automated} leveraged MIP to extract meaningful insights of lung nodules from 2D views (i.e., axial, coronal, and sagittal) and used multidimensional region-based fully convolutional neural network based CAD system for lung nodule detection and classification, respectively. Similarly, Zheng \textit{et al.} \cite{zheng2019automatic} attempted to follow the clinical workflow by MIP with different slab thicknesses (i.e., 5 mm, 10 mm, 15 mm) to train four different UNet architectures and merged their outputs for nodule candidates detection. Additionally, they also trained two different-sized DNNs for false positive reduction. Overall, they improve the performance,  however, the ensemble of six different networks makes the CADe system very complex. In contrast, we propose a single-stage system that exploits the 3D patch of CT scan along with MIP images to detect the lung nodule. 
We highlight the difference between our work with the existing work in Table \ref{litraturer}. Below is a summary of the main contributions of this study in contrast to the existing literature. 

\begin{enumerate}

\item None of the studies have used self-distillation for lung nodule detection. We design a novel single-stage multi-encoder-based CNN that utilizes the self-distillation mechanism and improves the performance compared to state-of-the-art single-stage as well as most dual-stage frameworks on the public LIDC-IDRI dataset, 


\item This study is the first to exploit the outputs from auxiliary branches of the same architecture to complement the primary network for false positive reduction.  
  
  \item While some studies used MIP with CT scan for lung nodule detection, however, no studies use bidirectional MIP. We propose to utilize bidirectional MIP images of three slab thicknesses, i.e., 3, 5, and 10 mm, for lung nodule detection, which improves the network's ability to distinguish nodules from non-nodular structures in the lung region. 
  

\end{enumerate}

\begin{figure*}[t]
\centering
\includegraphics[width=\textwidth]{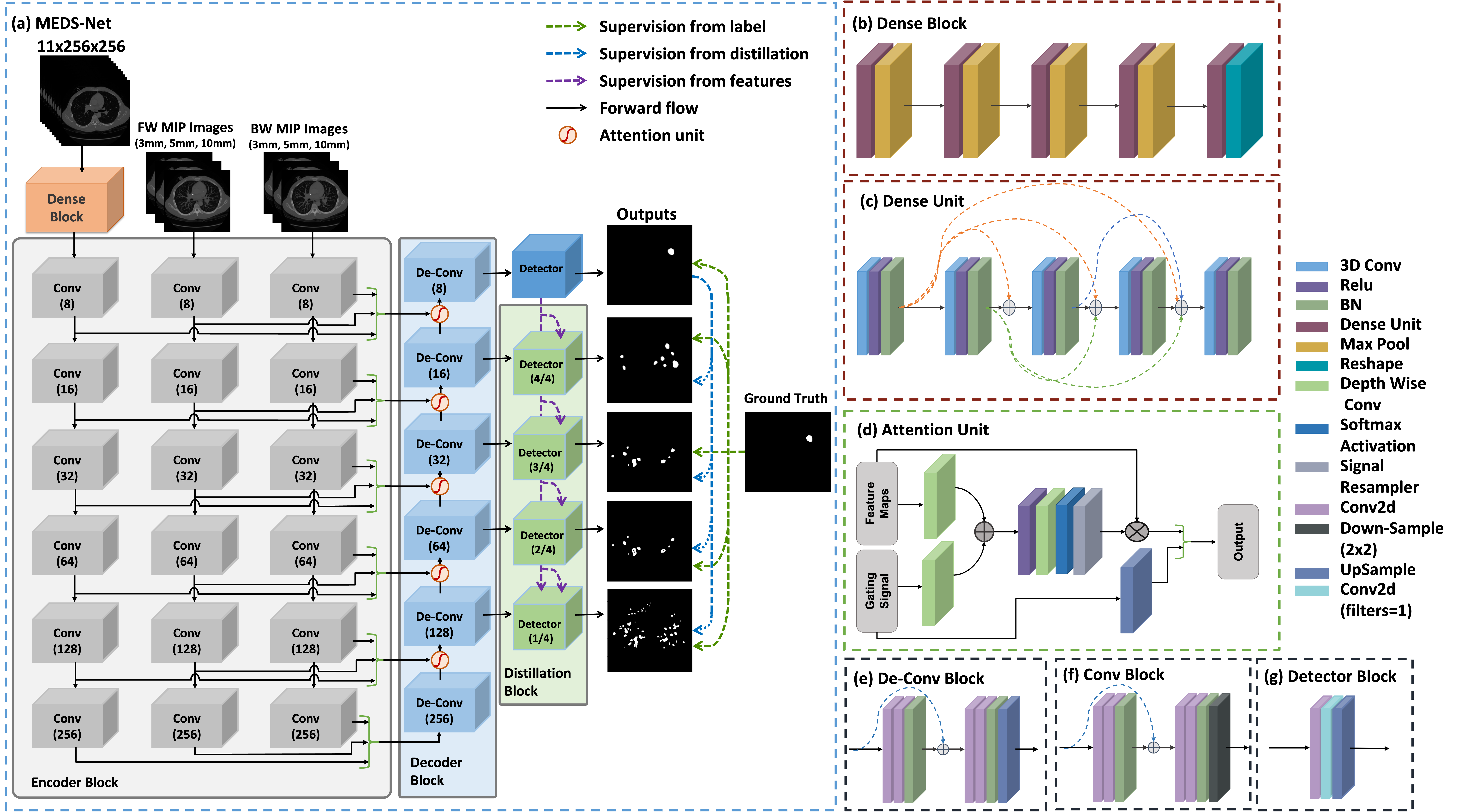}
\caption{The illustration of our proposed computer-aided detection system. (a) Multi-encoder based self-distilled network (MEDS-Net) which incorporates three different inputs, i.e., 3D sub-volume of CT scan, forward and backward MIP images of three slab thickness (i.e., 3, 5 and 10 mm). (b) The architecture of 3D dense block which transforms the 3D sub-volume of normalized thoracic CT scans into three-channel latent representation. (c) The architecture of dense unit utilized in the dense block. (d) Describing the architecture of attention unit used in residual connections from encoders to decoder block. Gating signal and feature maps come from the previous De-Conv block and the concatenated encoders features, respectively. (e) The architecture of convolutional block. (f) The architecture of de-convolutional block. (g) The architecture of detector block.}
\label{MEDSNet}
\end{figure*}

\section{Proposed Methodology}

In this section, we describe the proposed framework for lung nodule detection in abdominal CT scans, which emulate the clinical practices of radiologists who utilize maximum intensity projection (MIP) images \cite{wallis1989three}. Our proposed multi-encoder based self-distilled network (MEDS-Net) is designed to fully exploit the MIP images along with a 3D patch of CT scan as shown in Fig. \ref{MEDSNet}(a). To further extend the ability of MEDS-Net to detect the lung nodules, we enrich the inputs by extracting the MIP images from both forward and backward directions.
Most importantly, the proposed MEDS-Net leverages the self-distillation mechanism and auxiliary outputs to optimize the learning process and reduce false positives at the inference stage, respectively. The details of each component of the proposed framework have been described in the following subsections. 
\begin{figure}[!b]
\centering
\includegraphics[width=0.5\textwidth]{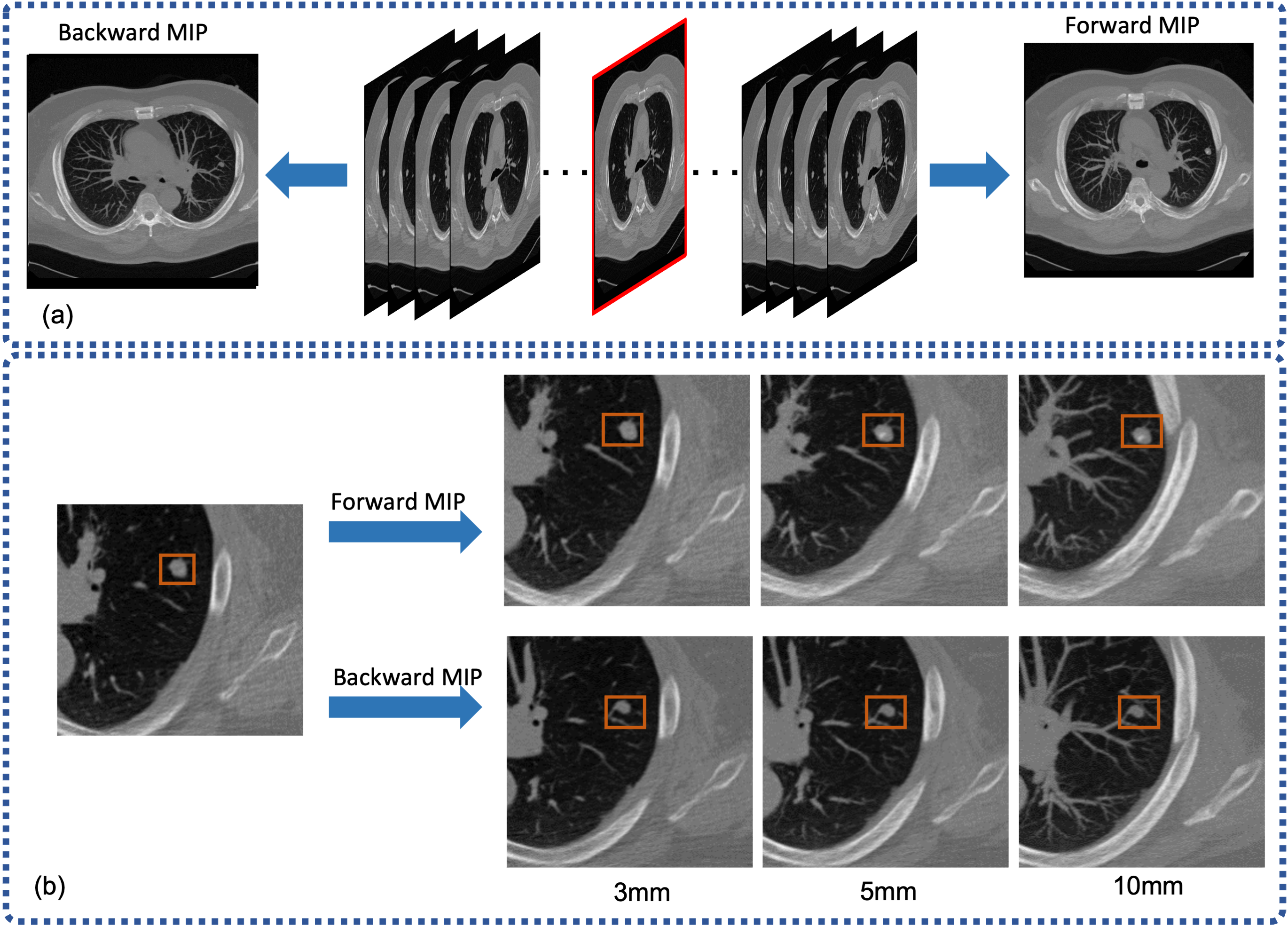}
\caption{Demonstration of bidirectional MIP images. (a) Process of generating the forward and backward MIP images. (b) Example of forward and backward MIP of thickness 3 mm, 5 mm, 10 mm (left to right). }
\label{mips_fw_bw}
\end{figure}

\subsection{Bidirectional Maximum Intensity Projection}
Maximum intensity projection (MIP) image consists of maximum values, which come across the plane of projection, i.e., the z-axis. MIP images of thoracic CT scans have proven extremely effective for lung nodule detection \cite{gruden2002incremental}; therefore, it has been utilized by radiologists to quickly screen lung nodules. MIP images can be generated of various slab thicknesses; however, research has shown that slab thicknesses up to 10 mm are the most effective ones, and increasing the slab thickness further reduces the detection sensitivity \cite{zheng2020deep}. Subsequently, in this work, we utilize MIP images with three slab thicknesses, i.e., 3 mm, 5 mm, and 10 mm. The generation of a MIP image can be described as follows:

\begin{equation}
\label{mip_eq}
I_{x,y} = \max _{\Delta z} D_{x,y,\Delta z}
\end{equation}
where, $I_{x,y}$ and $D_{x,y,\Delta z}$ represent the generated MIP image and sub-volume or slab of CT scan, respectively. Here, $\Delta z$ denotes the slab thickness, which in our case is set to 3, 5, and 10 mm. Traditionally, MIP is taken in one direction, which helps to distinguish nodules from the vessels penetrating in one direction only. In this work, we take it one step further to improve the visibility of nodules; bidirectional MIP images are extracted to improve the nodules' visibility. The process of generating the bidirectional MIP images, i.e., forward and backward directions, with the slab thicknesses of 3 mm, 5 mm, and 10 mm, has been demonstrated in Fig. \ref{mips_fw_bw}. We also normalize the slice thickness along the z-axis of the raw CT scan to 1 mm while keeping the original pixel spacing in xy-plane for 3D patch-based input.

\subsection{Multi-Encoder based Self Distilled Network}
The proposed multi-encoder based self distilled network (MEDS-Net) is an end-to-end lung nodule detection framework which is demonstrated in Fig. \ref{MEDSNet}(a) and summarized as follows: (1) first 3D slab consisting of a total of 11 slices (i.e., $11 \times 256 \times 256$) of 1 mm thickness is fed to dense block which squeeze the representation into three channels by using dense units; (2) the compressed features from dense block along with forward and backward MIP images are inputted into encoder block which contains three deep encoders to extract the meaningful features from each three-channel input; (3) the extracted features at various levels of three encoders are concatenated to combine with the corresponding decoder layers via attention units to decoder block; (4) decoder block utilizes the deconvolutional layers to up-sample the latent features at each stage that are utilized as gating signals to attention units. At various levels of the decoder, the features are reshaped to the higher dimensions to make same size as the input image to acquire the segmentation result; (5)  finally, outputs at four different levels have been up-sampled by employing deconvolutional layers in a deep supervision block. 
\subsubsection{Dense block}
To incorporate the spatial, sequential information crucial to detect the small nodules, we input a patch of the scan to the proposed architecture. Concretely, the five adjacent slices from both sides of the central slice, i.e., the 3D patch of depth $11 \times 256 \times256$, is extracted and inputted into 3D dense block, which is shown in Fig. \ref{MEDSNet}(b). 3D dense block takes 11 slices of 1 mm thickness as input and transforms into the meaningful latent representative of the same dimension as three MIP images, i.e., $3 \times256 \times256$. 3D dense block is composed of four sets of dense units followed by the max-pooling layers, and the fifth dense unit is followed by the reshaping layer, which squeezes the information to desired dimensions. Fig. \ref{MEDSNet}(c) shows the architecture of each dense unit which is composed of five sets of 3D convolution, Relu activation, and batch normalization. The output of each set is propagated to all the next sets using skip connections. After each dense unit features are down-sampled along the depth, subsequently, at the end of the dense block, we have three channels of feature maps. These feature maps are further fed into the encoder block to further refine the features pertaining to nodule detection.

\subsubsection{Encoder block}
In Fig. \ref{MEDSNet}(a), the overall structure of encoder block has been demonstrated, which includes three coding paths, namely, 3D patch and forward and backward MIPs encoders. All three encoders are kept identical and inspired by conventional UNet encoder \cite{ronneberger2015u}. Nevertheless, our encoders are deeper than the encoder in the standard UNet to effectively learn meaningful insights from lung structure which are crucial to distinguish the lung nodules. Each encoder consists of six Conv-Blocks which down-sample the input to lower dimensions while extracting the meaningful features. As shown in Fig. \ref{MEDSNet}(f), Conv-Block has two stages; the first stage has two 2D convolutional layers followed by a batch normalization layer, then a skip connection is used to combine the input with the output features of the first convolution stage. This type of short skip connection tends to stabilize gradient updates. In the second stage, the combined features are passed to the same set of layers as the first stage, followed by an additional downsample layer. The width of each of our branches is kept small to avoid overfitting. Our first Conv-Block contains eight feature maps that get doubled after every downsampling layer, and eventually, our final Conv-Block has 256 features. Finally, the output features from the second to sixth Conv-Blocks of all three encoders are concatenated to input into decoder block.


\subsubsection{Decoder block}
The proposed architecture leverages rich information extracted from three types of inputs, i.e., 3D sub-volume consisting of 1 mm thick CT slices, forward, and backward MIP images. This information is extracted from encoders at multiple scales by skip connections to feed into decoder block. In decoder block, we exploit the attention gates \cite{SCHLEMPER2019197} to enable the network to focus on the expected nodule area in naive features coming from encoder block while discarding the redundant ones. The attention units achieve this goal by performing element-wise multiplication of input feature-maps and attention coefficients to highlight important features as described in Fig. \ref{MEDSNet}(d). Overall, the decoder block consists of six deconvolution (DeConv) block which up-samples the input features while extracting useful information at different levels. Fig. \ref{MEDSNet}(e) describes the architecture of DeConv block, which consists of two stages. In the first stage, two convolution blocks are followed by a batch normalization layer. Input and output features of stage one are combined using a skip connection with the attention unit and forwarded to stage two. The second stage is similar to the first one, followed by the up-sampling layer. The decoder increases the dimension of features step by step till the final deconvolution block, which has the same feature dimension as the ground truth mask. Further, we extract the outputs of each DeConv-Block to feed forward to the distillation block, which utilizes these features to reconstruct the nodule masks.

\subsection{Self-Distillation Mechanism}
\label{sdm_section}
The proposed framework employs a self-distillation mechanism to improve the training process by using a self-distillation block as shown in Fig. \ref{MEDSNet}(a), which has been proven effective in optimizing the network's performance for classification tasks \cite{zhang2019your}. We extract the features from five DeConv-Blocks within the decoder block to pass into the self-distillation blocks, which consist of four auxiliary detectors and the main detector block. The main detector is connected with the deepest DeConv-Block in the decoder block, therefore, having access to enrich insights for lung nodule detection. Each detector is responsible for immediately up-scaling the features from decoder block to generate the lung nodule candidates. Each detector-block shares the same architecture demonstrated in Fig \ref{MEDSNet}(g). The first layer of the detector block is an up-sampling layer that directly up-scale the features to the same dimensions as inputs' height and width, i.e., $256\times256$. The second layer is a 2D convolution layer with Relu activation, and the final layer is a convolution layer of single filter and softmax activation. 

During the training process, all the auxiliary detectors are trained as student models via distillation from the main detector, which can be conceptually regarded as the teacher model. To improve the learning ability of student detectors in the proposed framework, we utilize three types of losses during the training processes:\\
\textbf{Loss 1:} It is the dice loss that is calculated with labels of training data to all the detector networks outputs. It provides equal opportunity to each detector network to leverage the supervision from actual labels for optimizing their weights.
\textbf{Loss 2:} KL (Kullback-Leibler) divergence loss under the main detector’s supervision, which acts as a teacher network. We utilize the outputs of each auxiliary detector, i.e., softmax output, and output of the main detector to compute the KL divergence. This loss helps distill the knowledge from the main detector to the auxiliary detectors, forcing them to learn to produce similar results as the deepest detector. \\
\textbf{Loss 3:} It is an L2 loss that is computed between the feature maps of the main detector network and each auxiliary detector. The loss enables the auxiliary detectors to distill the implicit knowledge in feature maps from the main detector, which induces the auxiliary detectors' feature maps to fit the feature maps of the main detector.



\subsection{False Positive Reduction using Auxiliary Detectors}
In contrast to previous studies, the false positive reduction is performed by a separate network; our framework exploits the auxiliary detectors to reduce the false positives at the time of inference. All the nodules detected by the main detector are considered nodule candidates, and auxiliary detectors complement the detection of the main detector to validate the nodule candidate. Firstly, our main detector provides all the nodule candidates, and we determine the 3D bounding box, which acts as region of the proposal ($RoP$) for each nodule candidate while incorporating all the nodular voxels. The similar $RoPs$ of the same dimensions and positions are extracted from the 3D outputs of all four auxiliary detectors to detect the false positives. Given that $n$ is the number of voxels in $RoP_i$ in the $ith$ detector, we can define true positive determination criteria as follows:    

\begin{equation}
\begin{aligned}
i s T P = t h r \left(\frac{1}{n k} \sum_{i=1}^k \sum_{j=1}^n RoP_{i}(j), \tau\right)
\end{aligned}
\end{equation}

\begin{equation}
\begin{aligned}
t h r(\theta, \tau) & =\left\{\begin{array}{l}
1, \theta>\tau \\
0, \text { otherwise }
\end{array}\right.
\end{aligned}
\end{equation}

In the above equations, $k$ is the number of auxiliary detectors, which in our case are four, and $\tau$ is the threshold of the nodule and non-nodule probabilities. While $thr(.)$ is a binarized function to binarize the
possibility value $\theta$, which is in the range of 0 to 1. We normalize the whole submissions of probabilities by $nk$ to limit the accumulation to 1. We apply the same criteria to all the nodule candidates after the inference. 
\begin{figure*}[t]
\centering
\includegraphics[width=0.9\textwidth]{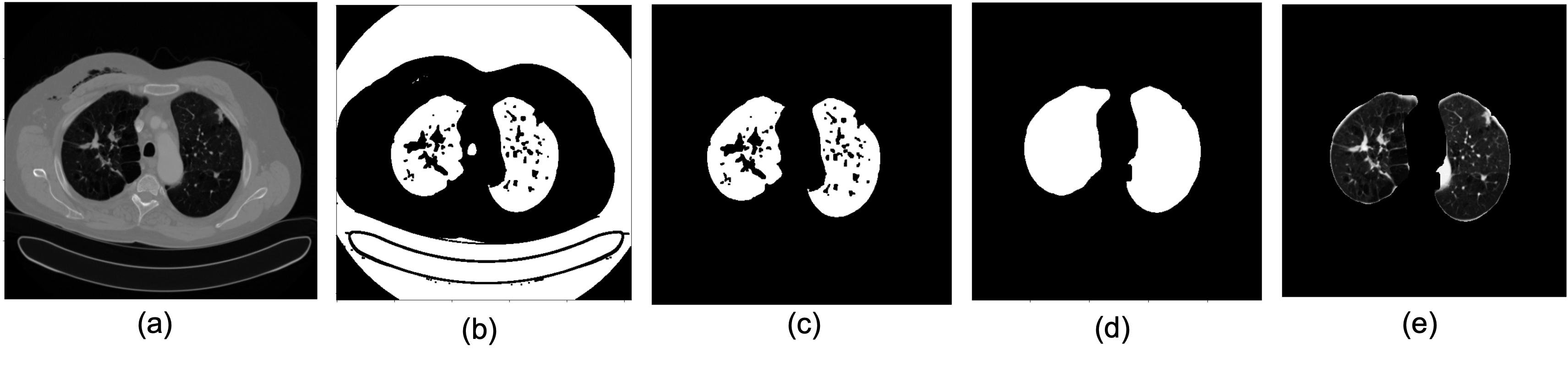}
\caption{Description of the steps involved in the lung parenchyma segmentation. (a) Raw thoracic CT image. (b) Mask of the air regions. (c) Mask after the removal of irrelevant objects other than the lung region. (d) Dilated mask of the lung parenchyma. (e) Image of segmented the lung parenchyma.}
\label{lungparcheyma}
\end{figure*}

\section{Experimental Setup}
\subsection{Datasets}

In this work, the lung image database consortium and image database resource initiative (LIDC/IDRI) dataset \cite{armato2011lung} has been utilized for all the experimentation. LIDC/IDRI dataset consists of 1018 thoracic computed tomography (CT) scans acquired from 1010 different patients, which are annotated to facilitate computer-aided systems on the assessment of lung nodule detection, classification and quantification. In the LIDC/IDRI dataset, slice thickness varies from 0.6 mm to 5.0 mm; however, we remove the scans with slice thickness greater than 2.5 mm as such low-resolution scans are not recommended for nodule screening \cite{kazerooni2014acr}. LIDC/IDRI dataset is annotated in two steps by four radiologists. Firstly, each radiologist annotated the nodules individually, and later all the radiologists jointly analyzed each annotation to reassess their annotations. For this study, we only utilized nodules having diameters equal to or greater than 3 mm due to their clinical significance \cite{national2011reduced}. In order to avoid possible outliers, we only considered the nodules accepted by at least 3 out of 4 radiologists as the reference standard. After applying the aforementioned criteria, 888  scans were left, which were utilized for all the experimentation. 

\subsection{Data Preprocessing}
We preprocess all the scans prior to feeding them into the proposed MEDS-Net. Our preprocessing consists of two stages; lung parenchyma segmentation and scan normalization. 

In the first stage, we automatically segment the lung parenchyma to exclude irrelevant regions such as clothes, machine objects, tissues, spines, or ribs in the scans. Fig. \ref{lungparcheyma} demonstrates the steps followed to perform lung segmentation. Initially, an air mask is created, and watershed algorithm \cite{angulo2007stochastic} is applied to remove the region other than the lung as shown in Fig. \ref{lungparcheyma}(b) and \ref{lungparcheyma}(c), respectively. In order to avoid missing wall-attached lesions that might be nodules, we keep more boundary information by binary morphology operations, i.e., closing and dilation. Finally, the binarized scan was masked with the original scan. 

In the second stage, we perform the scan normalization. Since the data set is collected from various CT scanners in the LIDC/IDRI dataset, we set the window level from -1000 HU to 400 HU and normalized images to the range between 0 and 1. We also normalized the slice thickness to 1 mm and cropped each slice equally from all sides while keeping the lung at the center for normalizing to $256 \times 256$ dimensions.

\subsection{Training Strategy}

We divide the whole dataset scans into eight subsets to perform 8-fold cross-validation. For each fold, training, validation, and test set split are set to 62.5\%, 12.5\%, and 25\% of the total scans, respectively. The batch size is set to 3 due to the memory limitations of the GPU. We used adam optimizer with an initial learning rate of 0.001 and first and second momentum of 0.9 for the decay of the learning rate. We use early stopping with patience of 10 epochs to avoid overfitting. 

This study exploits three types of losses to train the model as described in Section \ref{sdm_section}, which can be written as

\begin{equation}
Loss_{Total}= Loss_1 + Loss_2 + Loss_3
\end{equation}

To formulate these losses, we denote our main and auxiliary detectors by $\theta_{m}$ and $\theta_{i / k}$ detectors, respectively. Where, $i \in [\![1,k]\!] $ and $k$ is the number of auxiliary detectors.

\begin{equation}
Loss_1 = (1-\alpha) L_{DSC}(\theta_{m},y) + \sum_{i=1}^k L_{DSC}(\theta_{i / k},y)
\end{equation}

The first loss utilizes dice loss which is denoted as $L_{DSC}$ and can be defined as:

\begin{equation}
    L_{DSC} = \frac{2 . \mid x\cap y\mid}{\mid x \: \cup \: y\mid}
\end{equation}

The second loss Kullback-Leibler ($KL$) divergence between each auxiliary and main detector that can be written as:

\begin{equation}
Loss_2 = \alpha . \sum_{i=1}^k KL(\theta_{i / k},\theta_{m})
\end{equation}

The third supervision comes from the features of the main deepest detector to each auxiliary detector. Due to the difference in the size of features inputted to each detector, additional convolution layers in detector blocks are employed to align the dimensions prior to the extraction of feature maps to calculate the feature loss. Feature loss can be formulated as:

\begin{equation}
Loss_3 = \lambda . \sum_{i=1}^k \mid F_{i / k},F_{m}\mid
\end{equation}
where $F_{i/k}$ and $F_{m}$ represents the features from $\theta_{i / k}$ and $\theta_{m}$ detectors. While two hyper-parameters $\alpha$ and $\lambda$ are used to moderate the role of each loss.

\begin{table}[b]
\centering
\scriptsize
\caption{PERFORMANCE COMPARISON WITH OTHER COMPUTER-AIDED DETECTION SYSTEMS ON THE LIDC/IDRI DATASET AT NODULE CANDIDATE DETECTION STAGE}
\scalebox{1.1}{
\begin{tabular}{lccc}
\hline
\multicolumn{1}{c}{CADe systems} & \begin{tabular}[c]{@{}c@{}}Sensitivity \\ (\%)\end{tabular} & \begin{tabular}[c]{@{}c@{}}Total \\ number of \\ candidates\end{tabular} & \begin{tabular}[c]{@{}c@{}}Average \\number of \\ candidates \\per scan\end{tabular} \\ \hline
Zheng \textit{et al.} \cite{zheng2019automatic}           & 95.4                                                        & 18,116                                                                & 20.4                                                                            \\ \hline
Setio \textit{et al.} \cite{setio2015automatic}            & 31.8                                                          & 258,075                                                               & 290.6                                                                            \\ \hline
Yuan \textit{et al.} \cite{yuan2021pulmonary}            & 94                                                          & 44,627                                                                 & 50.25                                                                           \\ \hline
Pereira \textit{et al.} \cite{pereira2021classifier}         & 98.15                                                       & 44,111                                                                 & 49.67                                                                           \\ \hline
Wang \textit{et al.} \cite{wang2019pulmonary}            & 96.8                                                        & 53, 484                                                                & 60.2                                                                            \\ \hline
Zhang \textit{et al.} \cite{zhang2018nodule}           & 100                                                         & 45,939                                                                & 51.7                                                                            \\ \hline
Setio \textit{et al.} \cite{setio2017validation}          & 98.3                                                        & 754,975                                                               & 850.2                                                                           \\ \hline
Tarres \textit{et al.} \cite{lopez2015large}           & 76.8                                                        & 19,687                                                                & 22.2                                                                            \\ \hline
Our Method                      & 97.8                                                        & 19,190                                                                 & 21.61                                                                           \\ \hline
\end{tabular}
}
\label{candidates_detection_tab}
\end{table}

\begin{table*}[!h]
\centering
\scriptsize
\caption{COMPLETE PIPELINE PERFORMANCE COMPARISON WITH OTHER COMPUTER-AIDED DETECTION SYSTEMS ON THE LIDC/IDRI DATASET}
\scalebox{1.2}{
\begin{tabular}{lcccccccccc}
\hline
\multirow{2}{*}{CADe system} & \multirow{2}{*}{Year} & \multirow{2}{*}{Type of scheme} & \multicolumn{7}{c}{False positive per scan}                                                                         & \multirow{2}{*}{CPM} \\ \cline{4-10}
                            &              &         & 0.125         & 0.25           & 0.5            & 1              & 2              & 4              & 8              &                      \\ \hline
Setio \textit{et al.} \cite{setio2017validation}      & 2017  & Dual-stage                & 0.859         & \textbf{0.937} & \textbf{0.958} & \textbf{0.969} & \textbf{0.976} & \textbf{0.982} & \textbf{0.982} & \textbf{0.952}       \\ \hline
Wang \textit{et al.} \cite{wang2019pulmonary}        & 2018  & Dual-stage                & 0.788         & 0.847          & 0.895          & 0.934          & 0.952          & 0.959          & 0.963          & 0.903                \\ \hline
Zhang \textit{et al.} \cite{zhang2018nodule}       & 2018 & Dual-stage                 & \textbf{0.89} & 0.931          & 0.944          & 0.949          & 0.965          & 0.972          & 0.976          & 0.947                \\ \hline
Zheng \textit{et al.} \cite{zheng2019automatic}       & 2019  & Dual-stage                & 0.876         & 0.899          & 0.912          & 0.927          & 0.942          & 0.948          & 0.953          & 0.922                \\ \hline
Cao \textit{et al.} \cite{cao2020two}          & 2020  & Dual-stage                & 0.848         & 0.899          & 0.925          & 0.936          & 0.949          & 0.957          & 0.96           & 0.925                \\ \hline
Li \textit{et al.} \cite{li2020deepseed}          & 2020  & Single-stage                & 0.6           & 0.674          & 0.751          & 0.824          & 0.85           & 0.853          & 0.859          & 0.773                \\ \hline
Zhou \textit{et al.} \cite{zhou2022cascaded}        & 2022  & Dual-stage                & 0.742         & 0.84           & 0.8989         & 0.925          & 0.944          & 0.954          & 0.959          & 0.895                \\ \hline
Zhao \textit{et al.} \cite{zhao2022attentive}        & 2022  & Dual-stage                & 0.656         & 0.754          & 0.833          & 0.917          & 0.951          & 0.97           & 0.977          & 0.865                \\ \hline
Agnes \textit{et al.} \cite{agnes2022two}       & 2022     & Dual-stage             & 0.803         & 0.89           & 0.93           & 0.959          & 0.967          & 0.979          & 0.981          & 0.93                 \\ \hline
Zhu \textit{et al.} \cite{zhu2022channel}         & 2022   & Single-stage               & 0.782         & 0.834          & 0.893          & 0.917          & 0.932          & 0.952          & 0.956          & 0.895                \\ \hline
Mei \textit{et al.} \cite{mei2021sanet}        & 2022    & Dual-stage              & 0.712         & 0.802          & 0.865          & 0.901          & 0.937          & 0.946          & 0.955          & 0.874                \\ \hline
Guo \textit{et al.} \cite{guo2021msanet}         & 2022  & Single-stage                & 0.832         & 0.886          & 0.928          & 0.937          & 0.946          & 0.952          & 0.959          & 0.92                 \\ \hline
Our Framework               & 2022   & Single-stage               & 0.883         & 0.915          & 0.928          & 0.941          & 0.953          & 0.962          & 0.968          & 0.936                \\ \hline
\end{tabular}
}
\label{comp_SOTA}
\end{table*}

\section{Experimental Results and Discussion}

\subsection{Performance Analysis}
Typical CADe systems consist of two stages, i.e., candidate detection and false positive reduction, and mostly two separate networks are trained to achieve this goal. Nevertheless, in this study, the proposed framework combines both stages by employing self-distillation during the training and utilizing the auxiliary detectors for false positive reduction. For a comprehensive analysis of the proposed CADe system, we compare our performance with previously published methods for the candidate detection stage as well as after the false positive reduction stage. Below we compare the performance of the proposed MEDS-Net with multiple state-of-the-art studies. 

\subsubsection{Performance at nodule candidates detection stage}

All the nodules detected by the main, the deepest detector, of the proposed MEDS-Net are considered nodule candidates. The capability of nodule detection of the proposed framework has been compared with the recently published individual candidate detection systems. Table \ref{candidates_detection_tab} summarizes the results of candidate detection stages of previously published works. Setio \textit{et al.} \cite{setio2015automatic} and Tarres \textit{et al.} \cite{lopez2015large}, employed classical algorithms based on machine learning while Zheng \textit{et al.} \cite{zheng2019automatic}, Yuan \textit{et al.} \cite{yuan2021pulmonary}, Pereira \textit{et al.} \cite{pereira2021classifier}, Wang \textit{et al.} \cite{wang2019pulmonary}, Setio \textit{et al.} \cite{setio2017validation}, and our proposed framework utilizes the deep learning based methods. Zhang \textit{et al.} \cite{zhang2018nodule} developed a hybrid method by combing classical techniques with the deep learning-based algorithm. Overall the performance of deep learning-based methods is better than the conventional algorithms. Zhang \textit{et al.} \cite{zhang2018nodule}, Setio \textit{et al.} \cite{setio2015automatic} and Pereira \textit{et al.} \cite{pereira2021classifier} achieved better detection rate. However, these studies had improved the true positives at the cost of several-fold higher FPs.  Similar to our work, Zheng \textit{et al.} \cite{zheng2019automatic} exploited the MIP images of various thicknesses to achieve 95\% sensitivity with 20.1 candidates per scan. This shows the effeteness of MIP image for lung nodule detection. Although Zheng \textit{et al.} \cite{zheng2019automatic} trained four individual 2D networks with unidirectional MIP images, our framework with a single network outperforms in terms of sensitivity by exploiting bidirectional MIP images. The results demonstrate that incorporating bidirectional MIP images in a self-distillation network significantly improves the proposed architecture's nodule detection capability.

\subsubsection{Performance after false positives reduction} 
To provide a comparative analysis of the complete pipeline of the proposed framework, we chose the published approaches which used the competition performance metric (CPM) \cite{niemeijer2010combining} for evaluation on the LIDC/IDRI dataset. Table \ref{comp_SOTA} summarizes the performance of these techniques. Among all the listed methods, Setio \textit{et al.} \cite{setio2017validation} demonstrated the best performance, which is attributed to the combination of seven different nodule detection systems and five false positive reduction systems with varied architectures to detect the different types of nodules, such as subsolid, juxta-vascular, juxta-pleural nodules, etc. Nevertheless, the proposed scheme is designed for all types of nodules and consists of a single architecture. Our scheme exhibits comparable performance and even better sensitivity for the small value of FPs per scan (i.e., 0.125).
Similarly, Zhang \textit{et al.} \cite{zhang2018nodule} achieved better performance by maintaining 100\% sensitivity at the candidate detection stage with multi-scale LoG filters, which cost more FPs compared to our scheme. Our scheme exploits bidirectional MIPs with different levels of thickness to achieve comparable results. For the smaller value of FPs per scan, our sensitivity is only 0.7\% lesser, demonstrating the effectiveness of utilizing the auxiliary detectors for false positive reduction. Although Agnes \textit{et al.} \cite{agnes2022two} has a lower CPM score, their sensitivities are higher when the false positive rate is bigger than one. A possible explanation is that they had utilized a two-stage scheme in which they first exploited UNet+ architecture for nodule detection and employed LSTM-based architecture for false positive reduction, making the technique more complex. Similarly, Cao \textit{et al.} \cite{cao2020two} has shown better performance for larger values of false positive rate by using a two-stage strategy. Guo \textit{et al.} \cite{guo2021msanet} has obtained lesser CPM, and its performance degrades for a smaller value of false positives. Similarly, Li \textit{et al.} \cite{li2020deepseed}, Zhao \textit{et al.} \cite{zhao2022attentive}, and Mei \textit{et al.} \cite{mei2021sanet} obtained lower performance than the proposed scheme. Most importantly, our CADe system significantly outperforms single-stage systems, which proves that incorporating self-distillation and further utilizing the auxiliary detectors can significantly improve the lung detection rate efficiently.

\subsection{Ablation Study}
To demonstrate the effectiveness of each component incorporated into the proposed MEDS-Net architecture, we implemented various versions of proposed framework. We analyzed results after the nodule detection stage as well as after false positives reduction stage. The following subsections describe the details of these experiments and the corresponding results. 

\subsubsection{Analysis of candidate detection stage}
We initiate our evaluation by analyzing the contribution of each component of MEDS-Net for the lung nodule detection stage. For this purpose, we implemented eight downgraded versions of MEDS-Net, having different configurations and architecture, to evaluate their lung nodules detection performance. Table \ref{ablationTab} summarizes the results obtained from these models for lung nodules detection stage. First, to fairly analyze the effect of each input individually, models 1 to 3 utilize single encoder-based architecture which takes only one input among 3D sub-volume, forward, and backward MIP images, respectively. These networks are equipped with attention units and self-distillation mechanisms for lung nodule detection tasks. The results demonstrate that models 2 and 3 with MIP images of various thicknesses achieved better sensitivity with a lower number of candidates than the 3D input-based model. This proves that the incorporation of MIP images significantly improves the learning ability of networks which helps to better distinguish between nodule and non-nodular structures. Whereas, models 2 and 3 depict quite similar performance which shows that forward and backward MIP images provide the same level of information, i.e., unidirectional insights, to the network and their individual effect is the same. Models 4 and 5 are trained with two types of inputs by using dual-encoder based architecture having attention units and a self-distillation mechanism. Particularly, we combined 3D sub-volume with forward MIP images and forward MIP images with backward MIP images as inputs to models 4 and 5, respectively. It can be observed that combining the two types of inputs improves the sensitivity and also reduces the number of candidates per scan. However, model 4 performed slightly better than model 5 which can be attributed to the diverse type of inputs that enable the network to leverage the exceptional abilities of MIP images to provide deeper insights and 3D patch input that improves the localization of nodules.  
From models 6 to 9, we analyze the effectiveness of incorporating the attention units and self-distillation in our proposed architecture. For this purpose, we fixed our input pipeline and utilized multi-encoder-based architecture incorporating the 3D sub-volume and bidirectional (forward and backward) MIP images. Among the four versions, in model 6, we observe severe degradation in the sensitivity when attention units and self-distillation are removed. It also immensely increased the number of nodule candidates, which shows that the attention and self-distillation mechanism significantly contributes to reducing the number of false positives at the candidate detection stage. Concretely, the individual incorporation of attention units and self-distillation mechanism improve 3.94\% and 5.78\% sensitivity, respectively.
Most importantly, the self-distillation mechanism greatly reduces the number of candidates, improving the architecture's efficiency. Finally, the best performance is shown by model 9, which represents the proposed MEDS-Net architecture, having attention and self-distillation mechanism with three types of inputs. The results evidence that the combination of attention and self-distillation mechanism greatly increases our architecture's learning ability, enabling it to distinguish between the nodules and non-nodular structures in the complex lung region.  

\begin{figure}[b]
\centering
\centerline{\includegraphics[width=0.5\textwidth]{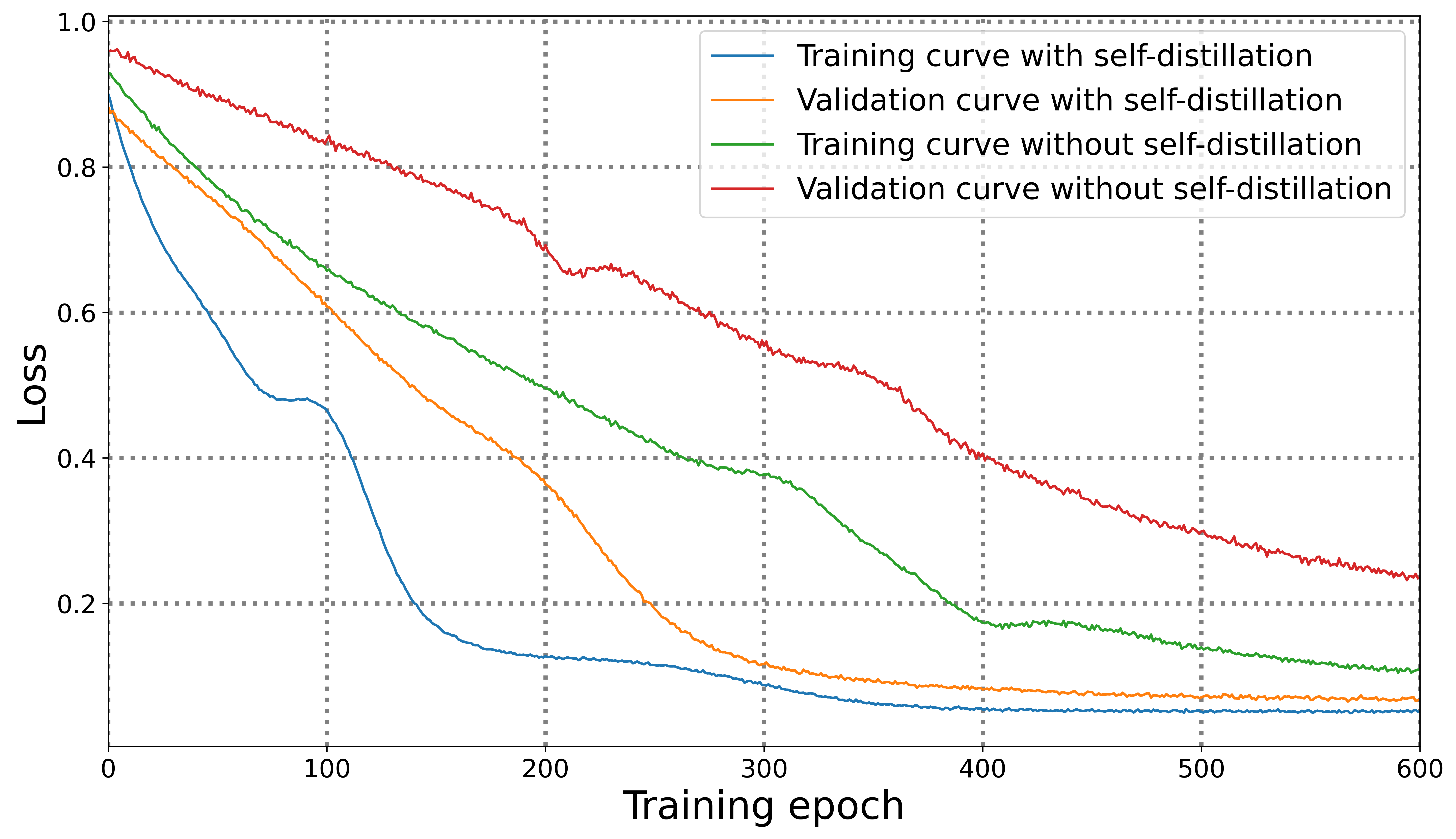}}
\caption{Comparison of learning curves of the training and validation curves of proposed multi-encoder based networks with and without self-distillation mechanism has been demonstrated.}
\label{tr_curves}
\end{figure}

To analyze the impact of self-distillation, we implemented the variant of the proposed architecture without self-distillation. Since MEDS-Net is a deeper network, it is critical to ensure that network is optimally trained without encountering vanishing gradient problem \cite{glorot2010understanding}. Subsequently, we analyzed the effect of self-distillation during the training process. The training and validation curves from the deepest detector of both variants have been shown in Fig. \ref{tr_curves}. Conceptually, self-distillation is an extended version of deep supervision that is proven to be effective in addressing the vanishing gradient problem for segmentation networks \cite{dou20163d}. The results show that the self-distillation mechanism in the proposed framework assists the training process to optimize network learning. It improves the convergence rate and helps optimize the learning by achieving lower training and validation losses. These improvements can be attributed to the supervision of the intermediate layers at multiple levels, which encourages the earlier layers to learn more meaningful and representative features for improved detection of lung nodules. 



We further analyze the performance of each detector, including main and auxiliary detectors, in the proposed MEDS-Net. Table \ref{detector_tab} summarized the results obtained from all the detectors, including the baseline implementation of multi-encoder based architecture without self-distillation. An ensemble result is obtained by including all the nodule candidates detected by each detector. The results of those detectors having performance lesser then the baseline model have been highlighted with red color. The results demonstrate that (i) all the networks have acquired significant performance improvement by exploiting the self-distillation mechanism. (ii) The proposed MEDS-Net has effectively detected almost all the nodules and achieved the sensitivity $99.41\%$ with its five detectors. (iii) The overall performance of deeper detectors is improved; however, several nodules missed by deeper networks have been detected by the less deep network, indicating that each detector has detected nodules independently. (iv) The self-distillation plays a crucial role in improving the deepest network, the main detector. The introduction of four auxiliary detectors improves the $4.08\%$ sensitivity of the main detector and significantly reduces the number of false positives per scan.

\begin{table*}
\centering
\caption{PERFORMANCE COMPARISON OF EACH DETECTOR OF PROPOSED MEDS-NET FOR LUNG NODULE DETECTION ON DIFFERENT SIZE OF NODULES}
\resizebox{\linewidth}{!}{%
\scriptsize
\begin{tabular}{lccccccc} 
\hline
\multicolumn{8}{l}{Total number of scans: 888} \\ 
\hline
\multicolumn{8}{l}{Total number of nodules: 1186 (3-10 mm: 905, 10-20 mm: 231, $\geq$20 mm: 50)} \\ 
\hline
\textbf{Feature Level} & \begin{tabular}[c]{@{}c@{}}\textbf{No. of detected }\\\textbf{nodules }\\\textbf{(3-10mm)}\end{tabular} & \begin{tabular}[c]{@{}c@{}}\textbf{No. of detected }\\\textbf{nodules }\\\textbf{(10-20mm)}\end{tabular} & \begin{tabular}[c]{@{}c@{}}\textbf{No. of detected }\\\textbf{nodules }\\\textbf{($\geq$20mm)}\end{tabular} & \begin{tabular}[c]{@{}c@{}}\textbf{Total No. of}\\\textbf{~detected }\\\textbf{nodules}\end{tabular} & \begin{tabular}[c]{@{}c@{}}\textbf{Sensitivity }\\\textbf{(\%)}\end{tabular} & \begin{tabular}[c]{@{}c@{}}\textbf{False Positives }\\\textbf{(FPs)}\end{tabular} & \multicolumn{1}{l}{\textbf{FPs per scan}} \\ 
\hline
\begin{tabular}[c]{@{}l@{}}Multi-Encoder Net \\(without Self-Distillation)\end{tabular} & 862 & 202 & 50 & 1114 & 93.93 & 28910 & 32.56 \\ 
\hline
Auxiliary Detector (1/4) & \textcolor{red}{841} & \textcolor{red}{195} & \textcolor{red}{46} & \textcolor{red}{1082} & \textcolor{red}{91.23} & \textcolor{red}{30226} & \textcolor{red}{34.04} \\ 
\hline
Auxiliary Detector (2/4) & \textcolor{red}{857} & \textcolor{red}{198} & \textcolor{red}{48} & \textcolor{red}{1103} & \textcolor{red}{93} & 26954 & 30.35 \\ 
\hline
Auxiliary Detector (3/4) & \textcolor{red}{856} & 205 & 50 & \textcolor{red}{1111} & \textcolor{red}{93.68} & 22047 & 24.83 \\ 
\hline
Auxiliary Detector (4/4) & 877 & 212 & 50 & 1139 & 96.04 & 19338 & 21.78 \\ 
\hline
Main Detector  & 889 & 221 & 50 & 1160 & 97.81 & 18004 & 20.27 \\ 
\hline
Ensemble & 901 & 228 & 50 & 1179 & 99.41 & 36780 & 41.42 \\ 
\hline
\end{tabular}
}
\label{detector_tab}
\end{table*}

\subsubsection{Analysis after false positive reduction}
\begin{figure}[!b]
\centering
\centerline{\includegraphics[width=0.5\textwidth]{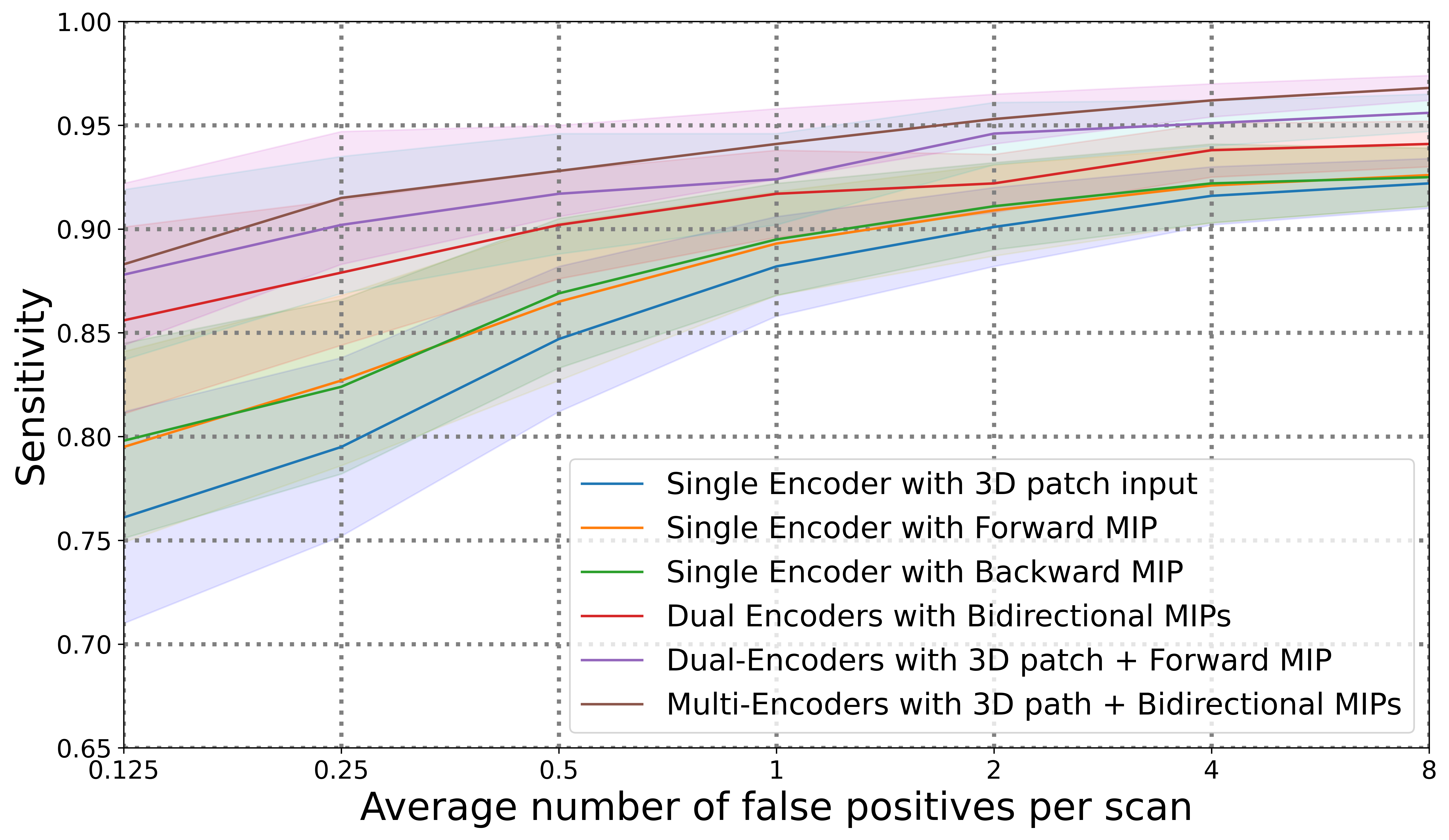}}
\caption{Free-response receiver operating characteristic (FROC) curves of various versions of proposed MEDS-Net architecture along with MEDS-Net results on the LIDC/IDRI database.}
\label{FROC_encoder_analysis}
\end{figure}

In this section, we analyze the effect of each input incorporated in the proposed MEDS-Net architecture on complete pipeline. Concretely, the proposed MEDS-Net exploits bidirectional MIP images along with the 3D patch to detect the lung nodules; therefore, it is important to analyze the impact of bidirectional MIP images after false positive reduction. We implemented single-encoder and dual-encoder versions of the proposed architecture with a complete pipeline that includes the false positive reduction stage. We implemented three single-encoder-based variants that took one input at a time and trained those models with 3D patch, forward, and backward image inputs individually. Similarly, we implemented two dual-encoder based variants, i.e., one with bidirectional MIP images inputs and another with 3D patch and forward MIP images. All the variants are implemented with self-distillation mechanism and followed by the same pipeline. To compare the performances of these variants models with proposed MEDS-Net for nodule detection, we analyze the free-response receiver operating characteristic (FROC) \cite{bandos2009area} curves of each variant shown in Fig. \ref{FROC_encoder_analysis}. The single-encoder-based architecture with 3D patch input has the lowest performance, and it becomes worse when we reduce the number of false positives per scan. Since the 3D patch of scan consists of 11 slices of 1 mm thickness (i.e., of depth 10 mm) which is insufficient to extract detailed insights of the nodule in the complex surroundings within the lung, it is challenging to detect nodules without false positives. The nodule bigger than 10 mm appear similar to the vessels, and small vessels have quite similar structural appearance to the nodule, which confuses the network and results in poor performance. Since conceptually, both directions of MIP images contain the same amount of information, subsequently, the single-encoder versions with forward and backward MIP images demonstrate similar; however, better performance compared to its 3D patch variant. MIP images based models are able to achieve almost 80\% sensitivity even with 0.125 FPs per scan which demonstrates the effectiveness of MIP images for lung nodule detection. 

We also compared dual-encoder based architectures, for which we implemented two combinations. At first, we utilized bidirectional MIP images and achieved significantly improved performance, which can be attributed to the fact that MIP images enhance the visibility of lung nodules, making it easier for the network to distinguish them from other anatomical structures. Most importantly, bidirectional MIP images cover a reasonable depth of scan, i.e., 30 mm, which enables the network to better explore the insights of each suspected nodule region. However, MIP images have certain limitations while representing the information as they rely on maximum intensities, suppressing the information in the intermediate intensities. Therefore, radiologists examine the nodules in the raw CT slices after the initial screen in MIP images. To this end, we also implemented a dual-encoder based variant with a 3D patch and forward MIP images. The results demonstrate that the combination of 1 mm thick slices with unidirectional MIP images are more effective than bidirectional MIPs as it equips the network with rich information which contains high level 3D aspect in the form of MIP as well as low level, 2D slice level information that helps to distinguish and precisely locate the small nodules. Finally, combining raw 3D patch of scan with bidirectional MIPs images in MEDS-Net depicts the optimal and consistent performance for lung nodule detection.

\section{Conclusion}
This work proposes a novel multi-encoder based self-distillation network (MEDS-Net) that leverages bidirectional maximum intensity projection (MIP) images for lung nodule detection system. Particularly, along with a 3D patch of the scan, the proposed framework utilizes forward and backward MIP images of 3, 5, and 10 mm thickness as input to improve the network learning. 
Most importantly, unlike conventional computer-aided detection (CADe) systems, the study exploits the auxiliary detectors to reduce the false positives, which avoids the extra computational costs caused by the separate false positive reduction networks. The key highlights are as follows:
\begin{enumerate}
    \item The framework has been comprehensively evaluated using the widely used LIDC/IDRI public dataset. The results using the proposed MEDS-Net showed that this network has excellent detection performance for the diverse nodule morphology and has the ability to distinguish false-positive nodules accurately. The CPM scores of the proposed method can be as high as 0.936, superior to the most advanced competitive networks.
    
    \item MEDS-Net exploits multi-scale features learning by using intermediate auxiliary outputs branches originating from various levels of decoder block to minimize the false positives.
    
  \item The experiments have shown that employment of self-distillation significantly improves the learning process by supervising the intermediate layers of the decoder block. Ablation experiments have also demonstrated that all the components in the proposed MEDS-Net are chosen carefully for effective lung nodule detection. 

    \end{enumerate}

Future works include extending the proposed CADe system while performing the accurate segmentation of lung nodules and their classification in terms of texture and malignancy levels.



\end{document}